# Si$_3$N$_4$ ring resonator-based microwave photonic notch filter with an ultrahigh peak rejection

David Marpaung[1,*], Blair Morrison[1], Ravi Pant[1], Chris Roeloffzen[2,3], Arne Leinse[4], Marcel Hoekman[4], Rene Heideman[4], and Benjamin J. Eggleton[1]

[1] *Centre for Ultrahigh bandwidth Devices for Optical Systems (CUDOS), the Institute of Photonics and Optical Sciences (IPOS), School of Physics, University of Sydney, NSW 2006, Australia*
[2] *Telecommunication Engineering group, University of Twente, PO Box 217, Enschede, 7500 AE, the Netherlands*
[3] *SATRAX BV, PO Box 456, Enschede, 7500 AL, the Netherlands*
[4] *LioniX BV, PO Box 456, Enschede, 7500 AL, the Netherlands*
[*] *d.marpaung@physics.usyd.edu.au*

**Abstract:** We report a simple technique in microwave photonic (MWP) signal processing that allows the use of an optical filter with a shallow notch to exhibit a microwave notch filter with anomalously high rejection level. We implement this technique using a low-loss, tunable Si$_3$N$_4$ optical ring resonator as the optical filter, and achieved an MWP notch filter with an ultra-high peak rejection > 60 dB, a tunable high resolution bandwidth of 247-840 MHz, and notch frequency tuning of 2-8 GHz. To our knowledge, this is a record combined peak rejection and resolution for an integrated MWP filter.

## 1. Introduction

Microwave filters are key signal processing components for selecting desired information signals or to remove unwanted portion in the radiofrequency (RF) spectrum, with applications in wireless communications, radar, and radio astronomy [1,2]. In recent years, microwave photonic (MWP) filters have shown competitive performance to their electrical counterparts, with distinct advantages in terms of immunity to electromagnetic interference (EMI), high bandwidth, large frequency tuning, and reconfigurability of the filter response [1-5]. Nevertheless, most high performance MWP filters have been demonstrated in optical fibers [1-4], which is not prone to integration. To be competitive with state-of-the-art RF filters, it is essential for MWP filters to adopt photonic integrated circuit (PIC) technology [5] that will lead to integration of multi-functionalities such as modulation, signal processing, and photodetection in one chip. Thus, a key challenge here is to achieve comparable performance with electrical RF filters while incorporating the PICs. This is not, by any means, trivial, since PICs insertion can compromise key filter performance such as resolution bandwidth, rejection level, or noise figure.

In this paper, we focus on microwave notch filters. Such a filter is crucial for removing interferers in dynamic, wideband radio systems, such as cognitive [6] or ultrawideband (UWB) [7] radios. These applications demand high resolution filtering (3-dB width of a few tens of MHz) with a very high notch peak rejection (>50 dB). State-of-the-art microwave notch filters have achieved these requirements [8-10], but they are limited in tuning and reconfigurability. Recently, several tunable integrated MWP notch filters have been reported with good performance [11-17], but only a few show comparable resolutions to RF filters. Filters based on stimulated Brillouin scattering (SBS) in a chalcogenide waveguide [16,18] have bandwidths the range of 30-140 MHz, while filters based on LiTaO$_3$ WGM resonators [17] have an ultra-narrow bandwidth of 10 MHz. But these filters are limited in peak rejection/suppression (20 dB in [16], and 45 dB in [17]). In these filters, to achieve peak rejection beyond 50 dB is very challenging; especially if one use a conventional modulation technique (i.e. single-sideband (SSB) modulation [15, 16]) to map the optical filter response to the RF domain. For active filters like SBS, high peak rejection requires very high pump power. In resonator-based filters, high peak rejection can be achieved at critical coupling [19]. But as, explained later, these filters suffer from a trade-off between resolution and peak rejection.

Here, we report a simple yet powerful technique in MWP signal processing that allows the use of an optical filter with a shallow notch to exhibit a microwave notch filter with anomalously high rejection level. We use the combination of an electro-optic modulator (EOM) and an optical filter to manipulate the phase and amplitude of optical sidebands for creating a signal cancellation within the RF notch filter response, thereby enhancing its peak rejection [20]. We implement this technique using a low-loss, tunable Si$_3$N$_4$ optical ring resonator as the optical filter, and achieved a microwave photonic notch filter with a peak rejection > 60 dB, a tunable high resolution bandwidth of 247-840 MHz, and notch frequency tuning of 2-8 GHz. To our knowledge, this is a record combined performance in terms of peak rejection and resolution for an integrated MWP filter. Since our technique is applicable to a wide range of optical filters, we predict that it will lead to creation of integrated MWP notch filters with superior performance compared to state-of-the-art electrical filters, when implemented with ultra-low loss resonators [21,22], or on-chip SBS devices [23,24].

## 2. Limitations of conventional notch filter

A schematic of an all-pass ring resonator [18, 25], with the definition of the cross-coupling amplitude coefficient, $\kappa$, is shown in Fig. 1a. By tuning $\kappa$, the magnitude and phase response of the ring can be reconfigured, as show in Fig. 1b. The simplest way to use the ring resonance as a tunable notch filter is by employing a single-sideband (SSB) modulation scheme, where the optical resonance is used to remove a portion of power in the optical sideband. This filtered optical spectrum is then mapped into the microwave spectrum via mixing with optical carrier during photodetection process. This mechanism is illustrated in in Fig. 1c.

From a notch filter perspective, tuning $\kappa$ will change the filter bandwidth (full-width at half maximum, FWHM), and peak rejection, as depicted in Fig. 1b. Given a ring resonator with a round trip length $L$, a round-trip loss $a$, and a self-coupling coefficient, $r$, where $r^2=1+\kappa^2$, the peak rejection at resonance is given by [19]

$$T_p = \frac{(a-r)^2}{1-2ra+(ra)^2} \qquad (1)$$

While the FWHM is given by

$$\text{FWHM} = \frac{(1-ra)\lambda_r^2}{\pi n_g L \sqrt{ra}}. \qquad (2)$$

Here $n_g$ is the group index and $\lambda_r$ is the resonance wavelength.

For a given resonator loss (fixed $a$), according to Eq. (1), the peak rejection of the filter is maximized when $r=a$, a condition known as critical coupling. On the other hand, according to

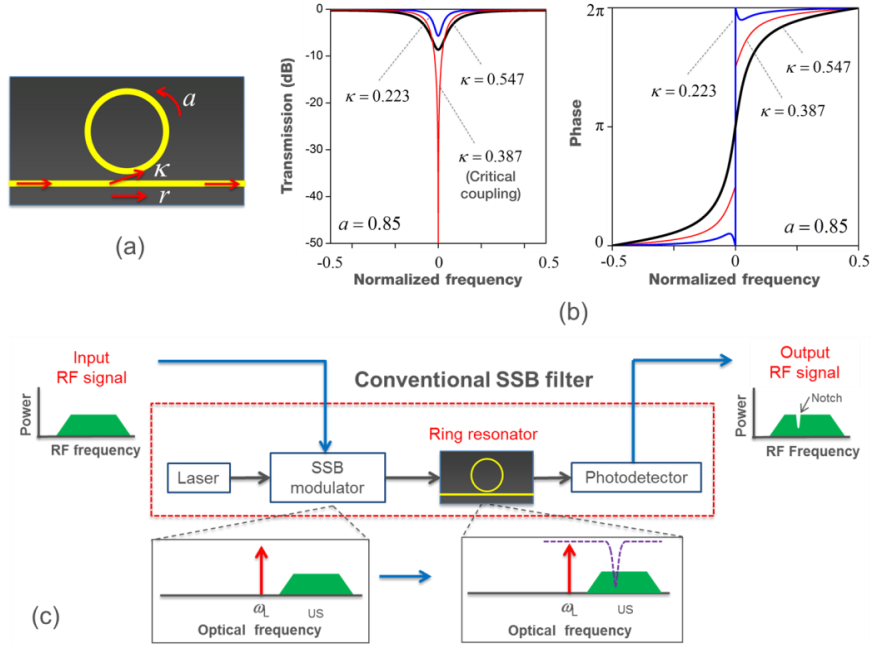

Fig. 1. (a) All pass ring resonator, with the definition of the tunable coupling coefficients and loss. (b) Simulated transmission and phase response of a ring resonator with varied coupling coefficient. (c) Schematic of a conventional single sideband (SSB) notch filter where the optical resonance of the ring resonator is mapped to the microwave frequency to exhibit a notch filter response.

Eq. (2), the FWHM is minimized when $r$=1. For a non-lossless resonator, $a$ is always smaller than unity ($a$<1), which means that in such a resonator one cannot simultaneously achieve minimum FWHM with maximized peak rejection. Thus, there is a trade-off between resolution and peak rejection for a conventional SSB-ring resonator microwave notch filter. For a low loss resonator, the FWHM at critical coupling is twice of the minimum FWHM.

This trade-off is clearly illustrated when we plot the peak rejection and FWHM for various values of $\kappa$ as shown in Fig. 2. In this simulation, we use parameters that correspond to the ring used in the experiment as the following: $\lambda_r$=1550 nm, $L$=8.783 mm, $n_g$=1.72, and $a$=0.974, which is equivalent to loss per unit length of 0.13 dB/cm. The minimum calculated FWHM of the ring was 166 MHz, which corresponds to an optical Q-factor [19] of 1.16 million. However, at this narrow bandwidth, the peak rejection is very low (< -0.5 dB). When $\kappa$ increases, the FHWM monotonically increases but the peak rejection improves until it reaches maximum at critical coupling ($\kappa$=0.226), where the FWHM has doubled to 332 MHz.

We performed experiments using a $Si_3N_4$ racetrack ring resonator with a radius of 125 μm and a total path length of 8.783 mm. The ring was fabricated using the TriPleX$^{TM}$ low-loss waveguide technology with double-stripe geometry [5, 26]. The group index of the optical waveguide was 1.72, leading to a free-spectral range of 19.8 GHz at 1550 nm, The propagation loss of the waveguide was previously reported in the order of 0.1 dB/cm [26]. Chromium heaters were deposited on top of the waveguide layer to enable thermo-optics tuning of the ring [11, 26]. The insertion loss of the ring away from the resonance wavelength was measured to be 13 dB, which was achieved by butt-coupling of standard single mode fibers. Using a setup as shown in Fig. 1c, we measured the down-converted ring magnitude response in the RF spectrum for various values of $\kappa$. For each measurement, we put the measured peak rejection along theoretical curve in Fig.2, and we compared the measured FWHM with the calculated one. There is a good agreement between the measured and the simulated values. The lowest FWHM from our measurements was 238 MHz, which correspond to an optical Q=814,000. At this bandwidth, the filter peak rejection was only -6.2 dB, which is not sufficient for a microwave notch filter.

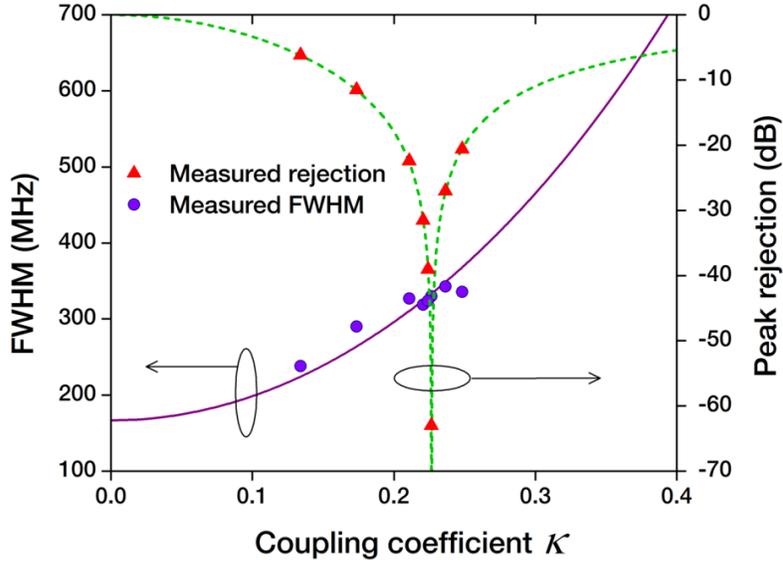

Fig. 2. Simulated and experimentally measured FWHM and peak rejection for the notch filter using $Si_3N_4$ ring resonator with parameters: $\lambda_r$=1550 nm, $L$=8.783 mm, $n_g$=1.72, and $a$=0.974.

### 3. Novel notch filter principle and experiment

The novel technique to significantly improve the peak rejection of the ring-based notch filter is shown in Fig. 3a. Instead of processing single sideband (SSB) signals, here we generate two sidebands with tunable amplitudes and phases [20]. The RF signal is encoded in the optical sidebands with unequal amplitudes and a phase difference, $\Delta\phi$, where $0 <$ modulo $(\Delta\phi) < \pi$, using an electro-optic modulator (EOM). We then exploit both amplitude and phase responses of the ring to equalize these sidebands amplitudes and to produce an anti-phase relation between them ($\Delta\phi = \pm\pi$), only in selected frequency region within the ring response. Upon photodetection, the beat signals generated from the mixing of the optical carrier and the two sidebands perfectly cancel at a specific microwave frequency, forming a notch with an anomalously high stopband rejection [20]. This is a photonic implementation of an absorptive band-stop filter in microwaves [8-10], that exhibits quasi-infinite rejection.

The optical sidebands with tunable phase difference and amplitude ratio can be generated using a dual-parallel Mach-Zehnder modulator (DPMZM) driven through a 90° (quadrature) RF hybrid coupler [27]. Such a modulator consists of two parallel MZMs nested in a larger MZ interferometer structure with a tunable phase-modulator (PM) in one of its arms. The DPMZM has three biases, two for the MZMs ($\theta_{MZ1}$ tand $\theta_{MZ2}$) and one for the PM ($\theta_{PM}$). We simulate the upper sideband (USB) and lower sideband (LSB) amplitudes and phase difference generated by the DPMZM at the output of our ring resonator ($a$=0.974, $r$=0.9873). When the MZM bias angles are adjusted to be $\theta_{MZ1}$ =1.7603$\pi$, $\theta_{MZ2}$ =0.5488$\pi$, and $\theta_{PM}$ =1.879$\pi$, the conditions of equal amplitude and opposite phase of USB and LSB are met at the center of the ring resonance. The sidebands amplitudes and phase difference are depicted in Fig. 3b and 3c, respectively.

The simulated normalized RF transmission generated from the mixing of USB and LSB with the optical carrier is shown in Fig. 3d, where it is compared to the filter response obtained from the conventional SSB scheme. Both filter responses have FWHM of 247 MHz, but the one generated with the novel phase cancellation scheme exhibited an ultra-high peak rejection of 70 dB, which is a 60 dB improvement from the peak rejection of the conventional

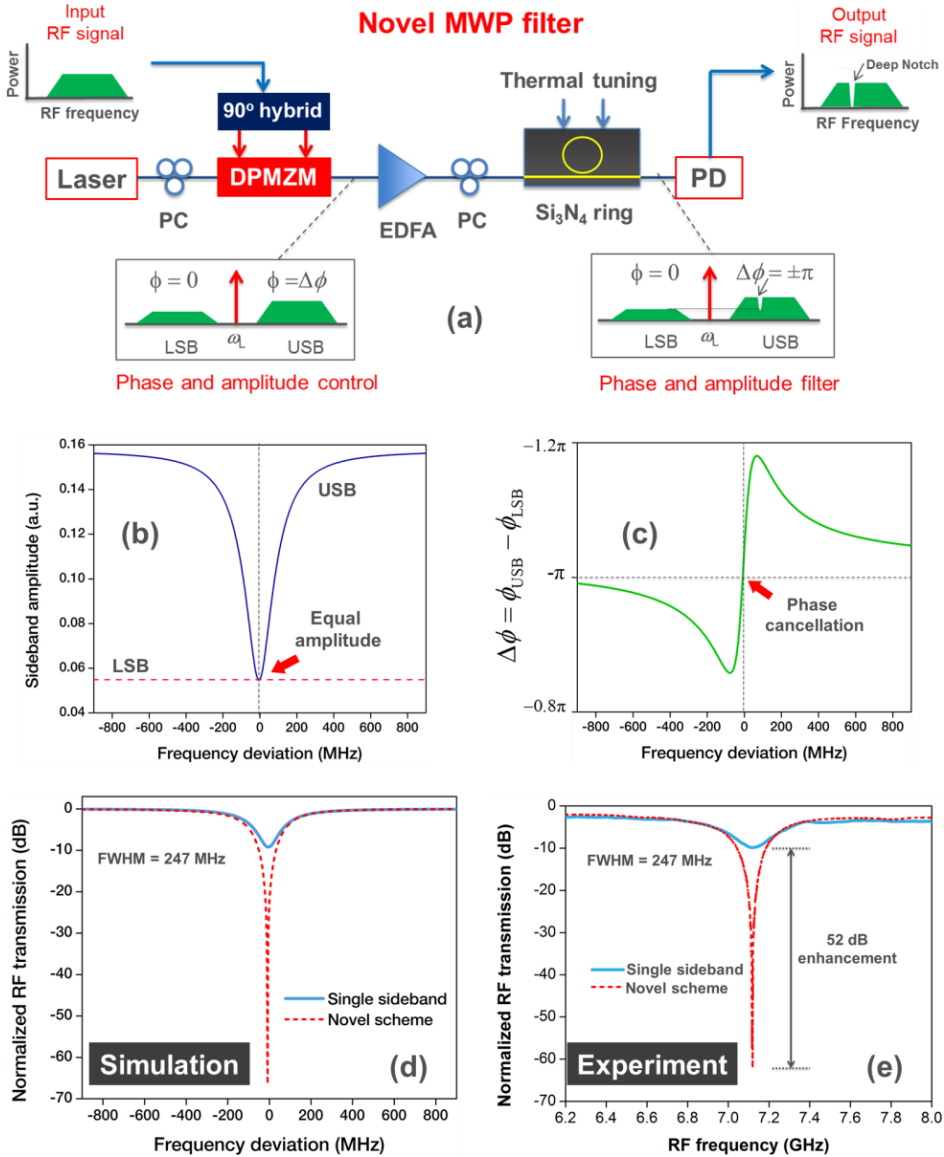

Fig. 3. (a) Topology of novel MWP notch filter that exploits phase and amplitude responses of ring resonator to create ultra-high peak rejection. DPMZ: dual-parallel Mach-Zehnder modulator, PD: photodetector. Sidebands amplitude (b) and phase difference (c) after modification in the DPMZ and ring resonator. Simulated (d) and experimentally measured (e) notch filter response of the novel MWP filter (dashed line) and conventional single-sideband filter (solid line).

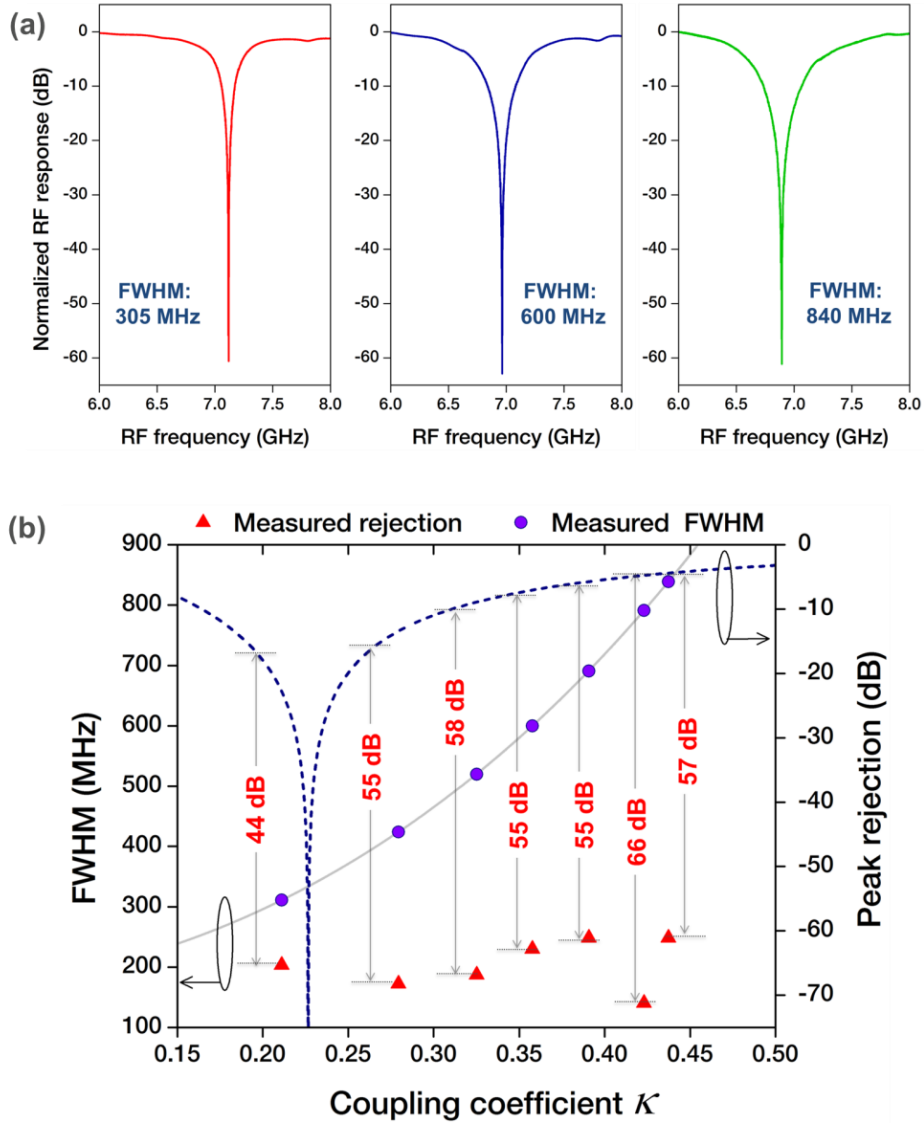

Fig. 4. (a) Experimentally measured tunable bandwidth filter responses with at least 60 dB peak rejection. (b) Experimentally measured peak rejection of the novel MWP filter (triangle) compared to the peak rejection of a conventional SSB case (dashed curve) for various filter FWHMs.

SSB filter.

To verify this simulation result, we performed experiments with the setup shown in Fig. 3a. We used a 1550 nm, 100 mW DFB laser (Teraxion PureSpectrum), a 20 GHz DPMZM (Covega Mach 40-086) driven via a quadrature hybrid coupler (1.7-36 GHz, Krytar), and a high-speed photodetector (PD, u2t XPDV2120). A low noise EDFA was used to increase the optical power before coupling to the ring resonator. We thermally tuned the coupler to the ring resonator to achieve FWHM of 247 MHz. The RF filter response was measured using a vector network analyzer (VNA). To maximize filter peak rejection, we controlled the three bias voltages of the DPMZM using a multi-channel programmable power supply with 1 mV accuracy (Hameg HM7044G). The measured filter response exhibited a peak rejection of 62 dB, as shown in Fig. 3e. This is a 52 dB improvement in peak rejection compared to the SSB filter.

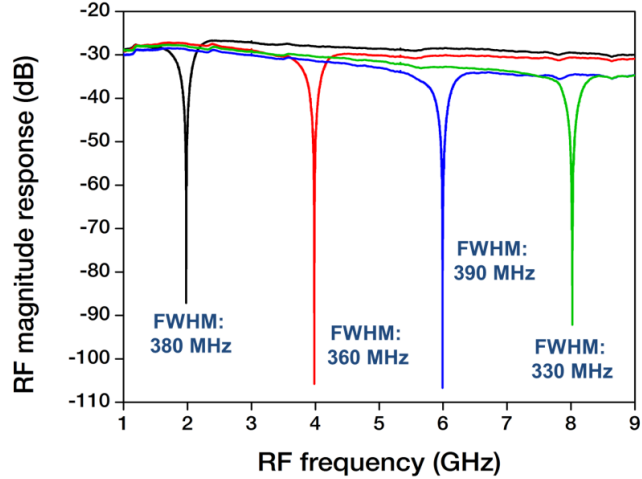

Fig. 5. Experimental results of frequency tuning of the MWP notch filter, preserving a narrow bandwidth of 350 MHz and ultrahigh rejection of >55 dB.

Furthermore, we demonstrate a bandwidth tunable notch filter by repeating the measurements for several values of FWHM from 305 MHz to 840 MHz, as shown in Fig. 4a. Each response exhibits a peak rejection of at least 60 dB. Fig. 4b highlights the main advantage of the proposed technique. Unlike the conventional SSB filter that suffers from trade-off between bandwidth and peak rejection when tuning the coupling of the ring (Fig. 2), the novel filter shows ultra-high peak rejection for all measured filter bandwidths. We compare the peak rejection of this filter (triangles in Fig. 4b) with the one of an SSB filter calculated from Eq. (2) (dashed line in Fig. 4b). The proposed filter shows impressive peak rejection enhancement, ranging from 44 dB up to 66 dB.

Finally, we demonstrated the frequency tuning of the filter by tuning the central frequency of our laser. Since the FSR of the ring is 19.8 GHz and we essentially use a double sideband modulated signal, the filter tuning range is half of the FSR, which is 9.9 GHz. The tuned filter response in the range of 1-9 GHz is depicted in Fig. 5. The insertion loss of the filter was in the order of -30 dB, which comprised the insertion loss of the DPMZ and the ring. The passband flatness degradation of the filter at higher frequency was attributed to the slope of the DPMZ frequency response. Over this tuning range, the filter showed peak rejection of at least 55 dB and maximum FWHM variation of 60 MHz.

## 4. Conclusions

A novel concept in sideband processing has been implemented using a low-loss $Si_3N_4$ ring resonator to achieve a tunable bandwidth microwave photonic notch filter with ultra-high peak rejection. Unlike previously reported RF notch filters based on ring resonator with multi-GHz isolation bandwidth and shallow peak rejection [13-15], our filter achieved a comparable resolution and peak rejection to state-of-the-art RF notch filters. The enabling signal processing concept reported here is applicable to a wide range of optical filters, such as fiber-Bragg gratings (FBGs) or active filters such as SBS, and will potentially lead to creation of very high performance integrated MWP notch filters in the future.